\documentclass[12pt]{iopart}
\usepackage[T1]{fontenc}
\usepackage[latin1]{inputenc}
\usepackage{subfigure}
\usepackage{floatflt}
\usepackage{graphicx}
\usepackage{setspace}

\makeatletter

\usepackage{iopams}
\makeatother

\def \dndy  {dN/dy}

\begin{document}

\title{ Probing collision dynamics at RHIC}

\author{Olga Barannikova \dag    \it{(for the STAR Collaboration})}

\address{\dag\ Department of Physics, Purdue University, West Lafayette, IN 47907, USA\\
barannik@physics.purdue.edu}

\begin{abstract}
Measurements of a variety of hadron species in pp and Au+Au collisions at 200 GeV are presented and studied within the framework of chemical and local kinetic equilibrium models. The extracted chemical and final kinetic freeze-out temperatures and collective flow velocities are  discussed as  function of centrality.  The results suggest that  Au+Au collisions of various centralities at RHIC always evolve toward the same temperature at chemical freeze-out, followed by cooling and expansion toward kinetic freeze-out. 

\end{abstract}


\section{Introduction}
\begin{floatingfigure}[r]{0.50\columnwidth}%
\begin{spacing}{0.8}
\hspace*{-0.4in}
\includegraphics*[
  width=0.50\columnwidth,
  keepaspectratio]{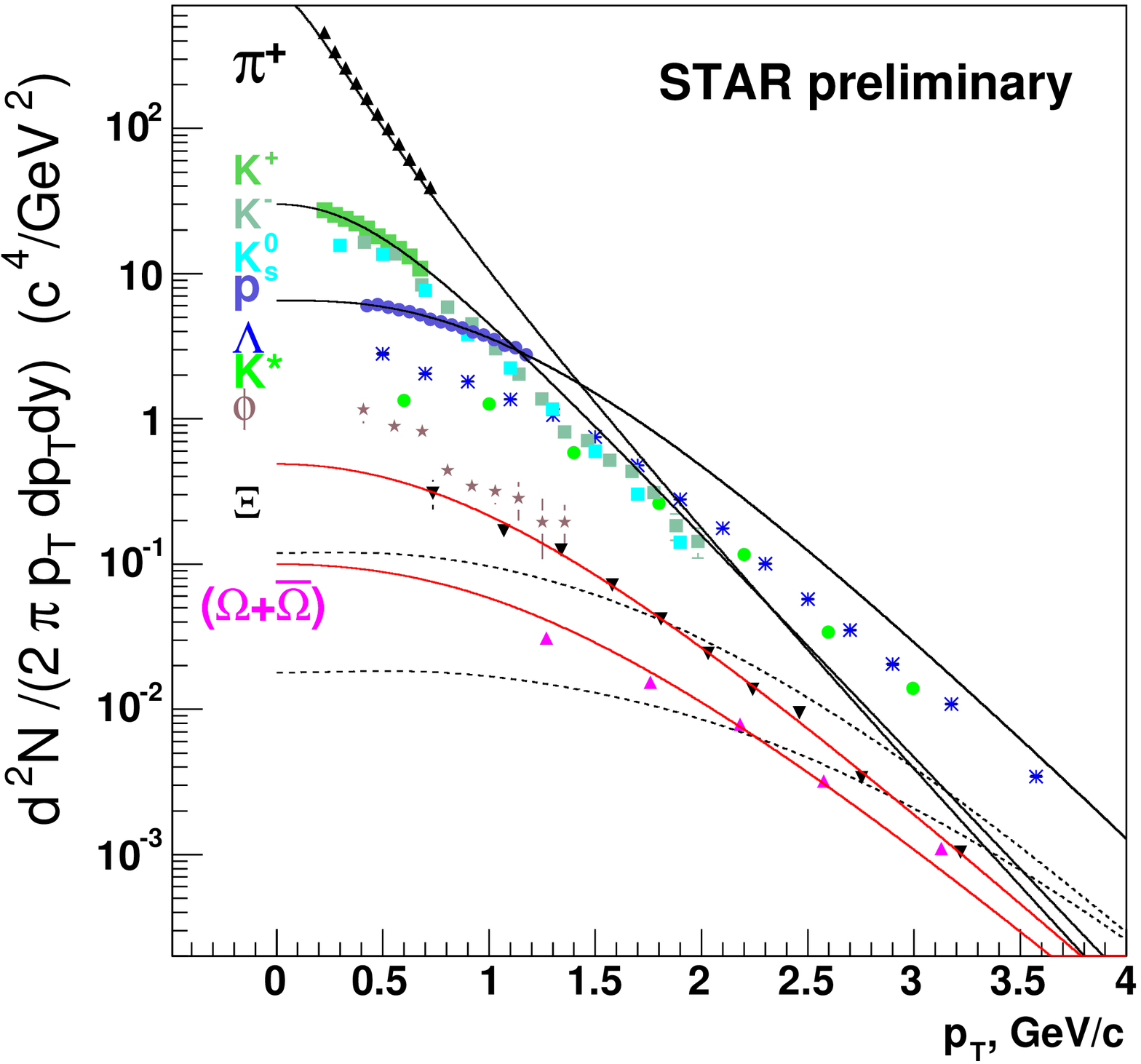}
\vspace*{-.25in}
\caption{ }
\noindent \footnotesize Identified particle spectra in 5\% most central Au+Au events measured by the STAR experiment. Black solid lines show blast-wave fit results for charged pions, kaons and protons. Dashed lines represent blast-wave prediction for multi-strange baryon spectra shapes based on freeze-out parameters of $\pi$, $K$, $p$. Red lines are blast-wave fits for $\Xi$ and $\Omega$ spectra.\smallskip{}
\end{spacing}
\end{floatingfigure}%

It is predicted that at sufficiently high energy densities matter is no longer confined to hadrons but exists in a deconfined state of quarks and gluons, the quark-gluon plasma (QGP). This new state of matter is expected to form in relativistic heavy ion collisions at RHIC. QGP signals may remain in the bulk properties of the collision at final state. For example, the final collective transverse radial flow, due to its cumulative nature, contains any collective flow generated at partonic stages, and particle yield ratios may reflect statistical nature of the hadronization process. We present here measurements of a variety of hadron species ($\pi^{\pm}$,  $K^{\pm}$, $K^0_s$, $K^*$, $\phi$, $p$, $\bar{p}$, $\Lambda$, $\bar{\Lambda}$,  $\Xi$, $\bar{\Xi}$, $\Omega+\bar{\Omega}$) in pp and Au+Au collisions at 200 GeV by the STAR experiment, and  a detailed investigation  of the final  hadronic state properties of such collisions in the framework of chemical and local kinetic equilibrium models.

\section{Data Analysis}

The STAR detector is described in \cite{star}. The main detector,  Time Projection Chamber (TPC), tracks charged particles in a uniform magnetic field of 0.5T. 
$\pi^{\pm}$, $K^{\pm}$, $p$, and $\bar{p}$ are identified be ionization energy loss; $K^0_s$, $K^*$, $\phi$, $\Lambda$, $\bar{\Lambda}$,  $\Xi$, $\bar{\Xi}$, and $\Omega+\bar{\Omega}$ are identified by their decay topology. 
Corrections for tracking inefficiency, detector acceptance as well as decay losses were applied. Pion spectra were corrected for weak decay feed-down, muon contamination and background pions; all other spectra include feed-down from weak decays. 
Additional corrections for  p+p spectra  were applied to account for inefficiency and false rate of the vertex finder.
Minimum bias Au+Au events were subdivided into  centrality classes based on measured number of charged particles within the pseudo-rapidity range $| \eta | < 0.5$. 

\begin{floatingfigure}[r]{0.50\columnwidth}%
\begin{spacing}{0.8}
\hspace*{-0.45in}
\includegraphics*[
  width=0.50\columnwidth,  
keepaspectratio]{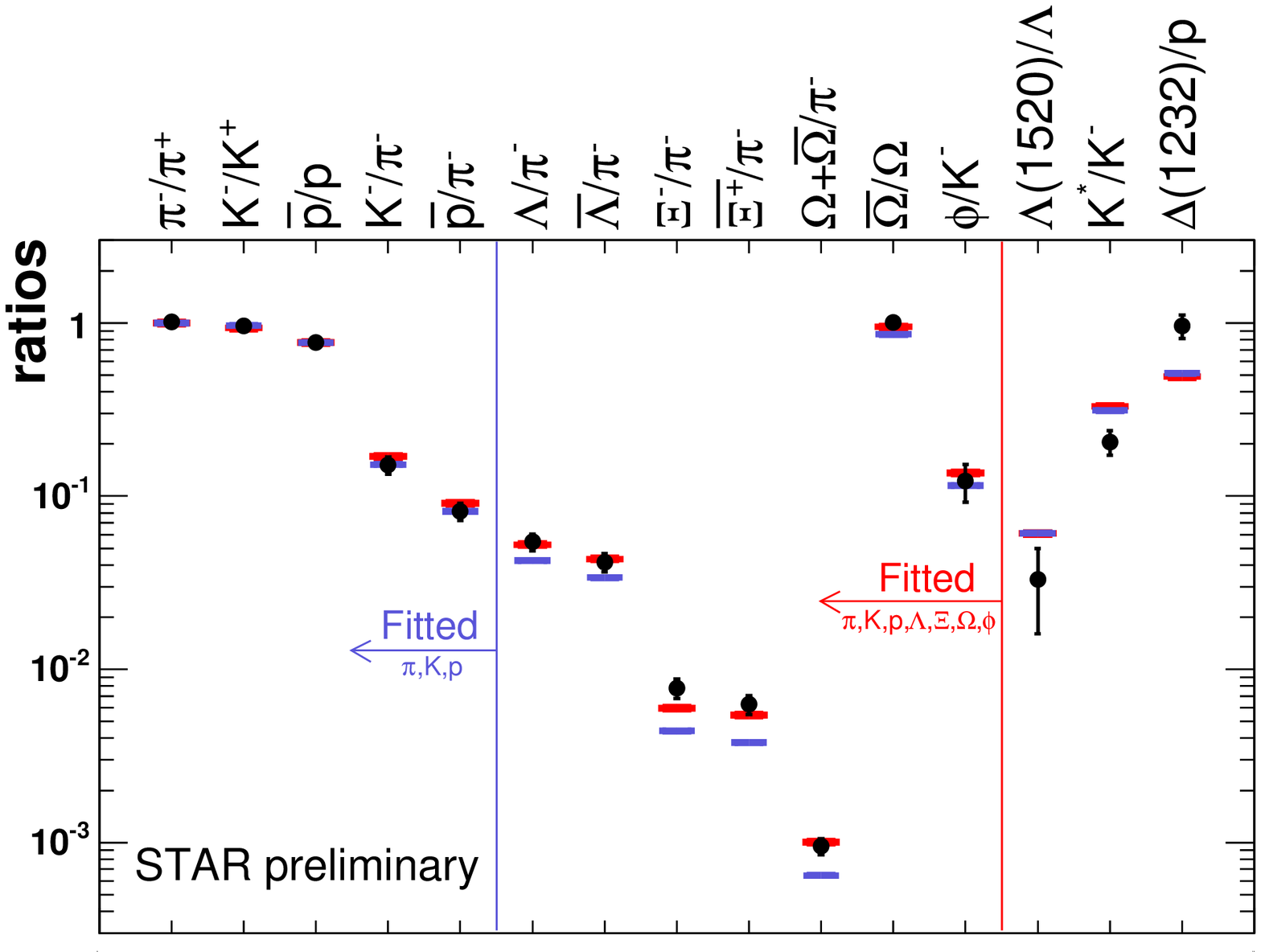}
\vspace*{-.05in}
\caption{ }
\noindent \footnotesize Measured particle ratios (symbols) and statistical model predictions (lines). Blue lines represent fit results based on $\pi$, $K$ and $p$ data. Red lines show results based on $\pi$, $K$, $p$, $\Lambda$, $\Xi$, $\Omega$ and $\phi$ data. \smallskip{}
\end{spacing}
\end{floatingfigure}%

Figure 1 shows comparison of the spectra for different particle types in central Au+Au collisions. Heavier particles show strong flattening of the spectra,  consistent with radial flow  resulting from high pressure gradient in the reaction zone. 
We fit $\pi^{\pm}$, $K^{\pm}$, $p$, and $\bar{p}$  spectra with a hydrodynamically motivated model with common flow velocity (also known as 'Blast-wave' model) with kinetic freeze-out temperature $T_{kin}$ and mean flow velocity  $\langle \beta \rangle$ as fit parameters \cite{BW}.  Fit results are  shown in Figure 1, demonstrating good description of $\pi$, $K$, $p$ spectra by such a fit (solid lines). Dashed lines shown in Figure 1 illustrate that blast-wave model fit to the common particle spectra fails to reproduce spectra shape of rare particles. 

Blast-wave fits for rare particle spectra are shown as red lines in Figure 1. Extracted values of  $T_{kin}$ and  $\langle \beta \rangle$ are given in Table 1. It has been suggested \cite{MSB} that multi-strange baryons freeze-out earlier than the common particles, possibly due to their smaller interaction cross-section with the bulk of the collision zone.

Integrated $\dndy$ values are extracted from the fit. Ratios obtained from those yields are fitted within the framework of statistical model with 4 parameters:  chemical freeze-out temperature ($T_{ch}$),  baryon and strangeness chemical potentials ($\mu_{B}$, $\mu_{s}$), and  strangeness suppression factor ($\gamma_{s}$) \cite{new2}. 
Points shown in Fig.2 represent measured STAR data for the most  central events, blue lines are statistical model results based on fit to the $\pi$, $K$, $p$ measurements. The obtained values of fit parameters are listed in Table 2 (first column).  As seen in the plot the relative abundances of strange particles are underestimated. Results of statistical model fit to all the ratios (data available are listed in the Table 3) are shown in red in Fig.2.  The thus obtained values of  $T_{ch}$, $\mu_{B}$, and $\mu_{s}$  agree with the  $\pi$, $K$, $p$ fit results, while strangeness suppression factor is larger, giving better description of strangeness production. The results are given in the last column of Table 2.
On the other hand  both fits fail to reproduce STAR resonance measurements, which may be due to rescattering/regeneration effects after chemical freeze-out.

Open symbols in Fig.3 summarize the centrality dependence of freeze-out parameters extracted from $\pi$, $K$, $p$ measurements: $T_{ch} \approx 160$~MeV and is independent of centrality, $T_{kin}$ decreases from $\approx 140$ MeV (peripheral) to $90$ MeV (central) and $\langle \beta \rangle$  and $\gamma_s$ increase with centrality~\cite{spectra}.  
The results suggest that  Au+Au collisions  of different initial conditions always evolve to the same chemical freeze-out condition, and then cool down further to a kinetic freeze-out dependent of centrality. 
The expansion of the system gives rise to collective flow. 
Significant drop in the temperature suggests relatively long time duration between two freeze-outs,  estimated to be at least  $6$~fm/$c$ for central events~\cite{spectra}.
%
\begin{floatingfigure}[r]{0.50\columnwidth}%
\begin{spacing}{0.8}
\hspace*{-0.4in}
\includegraphics*[%
   width=0.50\columnwidth,
   keepaspectratio]{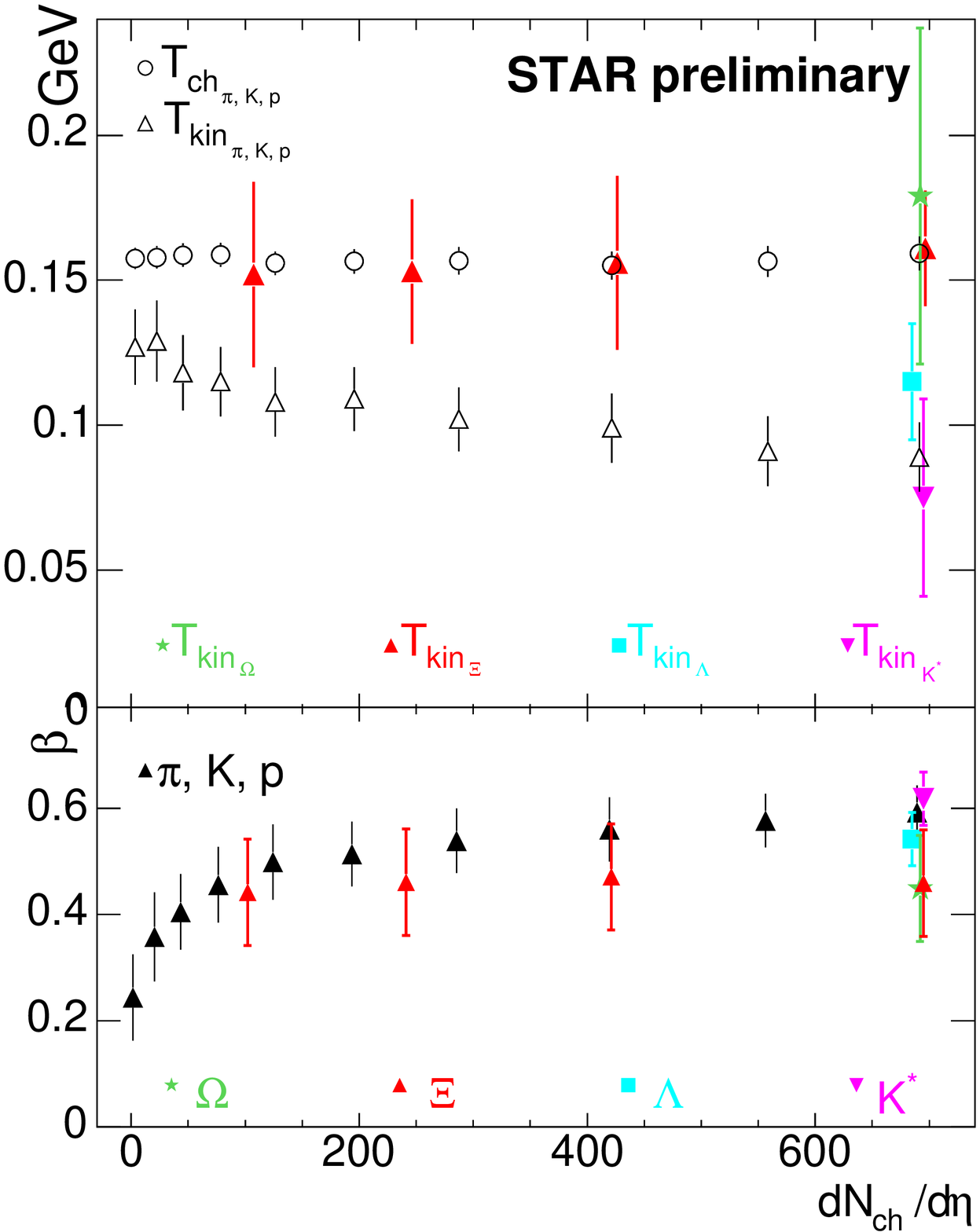}
\vspace*{-.25in}
\caption{ }
\noindent \footnotesize  Upper panel: chemical and kinetic  freeze-out temperatures from $\pi$, $K$, $p$ data; kinetic freeze-out temperatures for  $\Xi$, $\Omega$, $\Lambda$ and $K^*$ data separately. Lower panel: Mean transverse flow velocity for  $\pi$, $K$, $p$ (black), $\Xi$ (red), $\Omega$ (green), $\Lambda$ (blue), and $K^*$ (purple).
The fit results for $pp$ data are shown at $dN_{ch}/d\eta \sim 0$.
\smallskip{}
\end{spacing}
\end{floatingfigure}%
It also appears that the chemical freeze-out temperature from our fit is close to the Lattice QCD predicted \cite{karsch} critical temperature of $T_c \approx 170$~MeV (shown as a red line in Fig.3).

Figure 3 also shows the {\it kinetic} freeze-out parameters  from blast-wave fits to other particles.
The $\Xi$ results of those fits are shown as function of centrality (red symbols); the fit results of other particles are for central collisions.
The data, although with large uncertainties, seem to indicate sequential kinetic freeze-out of particles: $\Omega$ and $\Xi$ decouple from the system right after chemical freeze-out, followed by other particles ($\Lambda$, $K^*$, $\pi$, $K$, $p$). 
%
Also it appears that the kinetic freeze-out of $\Xi^-$ and $\bar{\Xi}^+$ baryons coincide with chemical freeze-out for all centralities. It is therefore a tempting to conclude that the $\langle \beta \rangle = 0.45 \pm 0.1$~$c$ extracted from $\Xi^-$ and $\bar{\Xi}^+$  spectra characterize radial flow at chemical freeze-out, which is possibly build up prior hadronization in the earlier, probably partonic, stage of the collision.

\section{Conclusions}

In summary, STAR has measured a wide variety of hadron species produced in 200 GeV Au+Au collisions. The data are studied within the framework of chemical and local kinetic equilibrium models to  investigate final  hadronic state properties of heavy ion collisions at RHIC. The major results of our study are:

\begin{itemize}
\item{A chemical equilibrium model fit to the  ratios yields a $T_{ch}$  insensitive
to centrality. The estimated value of $Tch \approx 160 \pm 6$~MeV is close to the QCD
predicted phase transition temperature.}

\item{Kinetic freeze-out temperature of multi-strange baryons shows no
sensitivity to collision centrality and coincides with chemical freeze-out
temperature for all centralities. The obtained flow velocity $\langle \beta \rangle = 0.45 \pm 0.1$~$c$ may be considered as that at hadronization.}

\item{Blast wave model fit to the transverse mass spectra of different particle
species indicates likely sequential kinetic freeze-out ($\Omega$, $\Xi$,$\phi$ $\rightarrow$ $\Lambda$, $\pi$,$K$,$p$,$K^*$) and allows to estimate radial flow velocity at the chemical as well as
kinetic freeze-out.}

\item{Kinetic freeze-out parameters obtained from blast-wave fits of $\pi$,$K$,$p$
spectra are correlated with centrality: the more central the collision, the
lower extracted temperature and the higher collective radial flow velocity. For the 5\% most central collisions, $T_{kin} = 89 \pm 10$~MeV,  $\langle \beta \rangle = 0.59 \pm 0.05$~$c$.
The drop in temperature from $T_{ch}$ to $T_{kin}$ and the development of strong
radial flow suggest a significant expansion and long duration from chemical
to kinetic freeze-out, estimated to be at least $6$~fm/$c$ in central collisions. }

\end{itemize}

The results seem to suggest the following picture at RHIC: Au+Au collisions of various centralities, despite different initial conditions, always evolve toward the same chemical freeze-out temperature (or possibly hadronization); chemical freeze-out is followed by cooling and expansion, and a sequential decoupling of particles dictated by their hadronic cross-sections: multi-strange baryons decouple at chemical freeze-out, followed by other particles decoupling at later times.

\begin{table}
\caption{\label{label}Blast-wave model fit results for 5\% most central Au+Au collision. Flow profile parameter $n$ is free parameter for $\pi$, $K$,  $p$ fit. The resulting value of $n=0.82$ is fixed for rare particle fits.}
\begin{indented}
\item[]\begin{tabular}{@{}c|c|c}
\br
Particle  & $T_{kin}$ (MeV) & $\langle \beta \rangle$ ($c$)\\
\mr
$\pi$, $K$, $p$           &  $89 \pm 10$  & $0.59 \pm 0.05$ \\
$K^*$ 			  &  $75 \pm 35$  & $0.62 \pm 0.05$ \\	
$\Lambda$, $\bar{\Lambda}$& $115 \pm 20$  & $0.54 \pm 0.05$ \\
$\Xi^-$, $\bar{\Xi}^+$	  & $161 \pm 20$  & $0.46 \pm 0.10$ \\
$\Omega$, $\bar{\Omega}$  & $179 \pm 60$  & $0.45 \pm 0.10$ \\
\br
\end{tabular}
\end{indented}
\end{table}

\begin{table}
\caption{\label{label}Chemical freeze-out parameters from statistical model fits to the particle ratios measured by STAR in  central Au+Au collision.}
\begin{indented}
\item[]\begin{tabular}{@{}c|c|c}
\br
Parameters  & $\pi$,$K$,$p$ ratios fit  & $\pi$,$K$,$p$,$\Lambda$,$\phi$,$\Xi$,$\Omega$ ratios fit\\
\mr
$T_{ch}$ (MeV)   & $157 \pm 6$      & $160 \pm 5$     \\
$\mu_B $ (MeV)   & $22  \pm 4$      & $ 24 \pm 4$     \\
$\mu_s $ (MeV)   & $ 3.8\pm 2.6$    & $ 1.4 \pm 1.6$  \\
$\gamma_s $      & $ 0.86 \pm 0.11$ & $ 0.99 \pm 0.07$\\
\br
\end{tabular}
\end{indented}
\end{table}

\begin{table}
\caption{\label{label}Particle ratios (errors are systematic) and statistical model fit results for 5\% most central Au+Au collision. Ratios from the bottom portion of the table were not included in fit.}
\begin{indented}
\item[]\begin{tabular}{@{}c|c|c}
\br
Ratio  & STAR data & Model prediction \\
\mr
$\pi^-/\pi^+$           & $1.01   \pm 0.02      $ & $ 1.00$   \\
$K^-/K^+$              & $0.96   \pm 0.03      $ & $ 0.93$   \\
$\bar{p}/p$            & $0.77   \pm 0.04      $ & $ 0.77$   \\
$K^-/\pi^-$            & $0.15   \pm 0.02      $ & $ 0.17$   \\
$\bar{p}/\pi^-$        & $0.082  \pm 0.009     $ & $ 0.091$  \\
$\Lambda/\pi^-$        & $0.054  \pm 0.006     $ & $ 0.052$  \\
$\bar{\Lambda}/\pi^-$  & $0.041  \pm 0.005     $ & $ 0.043$  \\
$\Xi^-/\pi^-$          & $7.8\cdot 10^{-3}\pm 10^{-3}     $ & $ 6.0 \cdot 10^{-3}$\\
$\bar{\Xi}^+/\pi^-$    & $6.3 \cdot 10^{-3}\pm 8\cdot  10^{-4}     $ & $ 5.5\cdot  10^{-3}$ \\
$\Omega/\pi^-$         & $9.5 \cdot 10^{-4}\pm 10^{-4}     $ & $10.1 \cdot 10^{-4}$\\
$\bar{\Omega}/\Omega$  & $1.01   \pm 0.08      $ & $ 0.95$    \\
 \br
$\Delta(1232)/p$       & $0.24   \pm 0.04$ & $ 0.12$    \\
$\phi/K^-$             & $0.12   \pm 0.03 $ & $ 0.14$    \\
$K^*_0/K^-$            & $0.205  \pm 0.033$ & $ 0.329$   \\
$\Lambda(1520)/\Lambda$ & $0.033  \pm 0.017$ & $ 0.061$   \\
\br
\end{tabular}
\end{indented}
\end{table}

 \end{document}